\documentclass[iop]{emulateapj}

\usepackage{ulem}
\usepackage{bm}
\usepackage{color}
\usepackage{comment}
\usepackage{lineno}

\newcommand
\kms{\ifmmode{\rm km\thinspace s^{-1}}\else km\thinspace s$^{-1}$\fi}

\newcommand\hstar{HD\,284163}

\newcommand\actaa{Acta Astronomica}

\shortauthors{Torres}
\shorttitle{\hstar}

\begin{document}
\submitted{Accepted for publication in Monthly Notices of the Royal Astronomical Society}

\title{Orbits and Dynamical Masses for the Active Hyades Multiple System \hstar}

\author{
Guillermo Torres\altaffilmark{1}, 
Gail H.\ Schaefer\altaffilmark{2}, 
Robert P.\ Stefanik\altaffilmark{1}, 
David W.\ Latham\altaffilmark{1}, 
Jeremy Jones\altaffilmark{2}, 
Cyprien Lanthermann\altaffilmark{2}, 
John D.\ Monnier\altaffilmark{3}, 
Stefan Kraus\altaffilmark{4}, 
Narsireddy Anugu\altaffilmark{2}, 
Theo ten Brummelaar\altaffilmark{2}, 
Sorabh Chhabra\altaffilmark{4}, 
Isabelle Codron\altaffilmark{4}, 
Jacob Ennis\altaffilmark{3}, 
Tyler Gardner\altaffilmark{4}, 
Mayra Gutierrez\altaffilmark{3}, 
Noura Ibrahim\altaffilmark{3}, 
Aaron Labdon\altaffilmark{5}, 
Dan Mortimer\altaffilmark{4}, and 
Benjamin R.\ Setterholm\altaffilmark{3} 
}

\altaffiltext{1}{Center for Astrophysics $\vert$ Harvard \&
  Smithsonian, 60 Garden St., Cambridge, MA 02138, USA;
  gtorres@cfa.harvard.edu}

\altaffiltext{2}{The CHARA Array of Georgia State University, Mount Wilson Observatory, Mount Wilson, CA 91203, USA}

\altaffiltext{3}{Astronomy Department, University of Michigan, Ann Arbor, MI 48109, USA}

\altaffiltext{4}{Astrophysics Group, Department of Physics \& Astronomy, University of Exeter, Stocker Road, Exeter, EX4 4QL, UK}

\altaffiltext{5}{European Southern Observatory, Casilla 19001, Santiago 19, Chile}

\begin{abstract}
We report near-infrared long-baseline interferometric observations of the Hyades
multiple system \hstar, made with the CHARA array, as well as almost 43~yr of
high-resolution spectroscopic monitoring at the CfA. Both types of observations
resolve the 2.39~d inner binary, and also an outer companion in a 43.1~yr orbit.
Our observations, combined with others from the literature, allow us to
solve for the 3D inner and outer orbits, which are found to be at nearly
right angles to each other. We determine the dynamical masses of the
three stars (good to better than 1.4\% for the inner pair), as well as
the orbital parallax. The secondary component ($0.5245 \pm 0.0047~M_{\sun}$) is now
the lowest mass star with a dynamical mass measurement in the cluster.
A comparison of these measurements with current stellar evolution models
for the age and metallicity of the Hyades shows good agreement. All
three stars display significant levels of chromospheric activity,
consistent with the classification of \hstar\ as an RS~CVn object.
We present evidence that a more distant fourth star is physically
associated, making this a hierarchical quadruple system.
\end{abstract}


\section{Introduction}
\label{sec:introduction}

The Hyades cluster has a long history of astrometric, photometric, and
spectroscopic observations dating back more than a century. 
{At a mean distance of only $\sim$45~pc, this collection of roughly 800
members \citep[e.g.,][]{Brandner:2023}, all of the same age and
chemical composition \cite[$\sim$750~Myr,
${\rm [Fe/H]} = +0.18$;][]{Brandt:2015, Dutra-Ferreira:2016}, has
served as a valuable laboratory for astrophysics. It is also} rich
in binary and multiple systems \citep[see, e.g.,][and references
therein]{Griffin:1985, Griffin:1988, Griffin:2012}, and yet
relatively few of them have had their {most basic property ---their masses---} determined
dynamically. 
For the most recent determinations
and a summary of earlier estimates, we refer the reader to the work
by \cite{Torres:2019} and \cite{Brogaard:2021}.
{Knowledge of the mass constrains models of stellar evolution
in ways that complement the information that can be obtained
perhaps more easily, such as luminosities and temperatures,
and permits valuable tests of theory. When in a cluster
of known age and composition, such as the Hyades, the constraint
is strengthened because there are then fewer free parameters that
one can adjust in the models in order to match the observations.
This is especially
interesting for compositions significantly different from solar,
as is the case for the Hyades.

Not all available mass measurements in the Hyades}
reach precisions at the level of 1--3\%, which
are most useful for a comparison with stellar models
\citep[see, e.g.,][]{Torres:2010}. Some are considerably worse,
with errors up to 15\%. {Furthermore, only well detached binaries
with no prior history of mass exchange are suitable for such tests.
We count seven suitable systems to date, excluding the case of
V471~Tau, which is a post-common envelope eclipsing binary composed of
a K dwarf and a DA white dwarf \citep{Muirhead:2022}. That system is
not representative of single-star evolution because of the
past interaction of the components.}

This paper reports precise mass measurements for {another} system in the
Hyades, \hstar\ (BD+23\,635, Pels\,20, V1136\,Tau; $V = 9.41$), as
part of an ongoing program designed to significantly enlarge the
sample of such determinations in the cluster. \hstar\ is classified as
a K0 dwarf, and was first recognised as a Hyades member by
\cite{Luyten:1971} and \cite{Pels:1975}. It has long been thought to
be a binary from its location well above the single-star main sequence
of the Hyades \citep{Upgren:1977}. It was found to be a single-lined spectroscopic
binary by \cite{Griffin:1981}, who established the period to be 2.39~days.
However, despite the $\sim$0.8~mag excess in brightness compared
to a single star of the same colour, those authors found only subtle and indirect
evidence of a companion from slight distortions in the radial
velocities obtained near conjunction. This was somewhat surprising,
as a companion raising the total brightness by as much as 0.8~mag should be rather obvious.
A note added in proof to
the \cite{Griffin:1981} paper finally showed this companion more clearly in a pair of
spectra obtained later, but pointed out the unexpected fact that its
velocity was stationary, indicating it is not the secondary in the
2.39~day orbit. They then remarked that the object ``...might well
repay careful examination, either directly or by interferometric or
occultation studies, for visual duplicity''. This suspicion proved
correct, and a few years later speckle interferometric observations by
\cite{McAlister:1987} revealed a visual companion at a separation of
about 0\farcs14, which we now identify with the stationary signal
found by \cite{Griffin:1981}. 
Subsequent speckle measurements by
others showed this wide companion to be in a $\sim$40 yr orbit around the
inner pair \citep{Mason:2010}, making the \hstar\ system a
hierarchical triple. 
Third components in close spectroscopic
binaries are not unexpected, as \cite{Tokovinin:2006} have shown that
more than 95\% of systems with periods under 3 days are attended by
a more distant companion.
The elusive secondary of the 2.39~day binary, an
M dwarf several magnitudes fainter than the primary, was finally
detected in near infrared observations by
\cite{Bender:2008}.\footnote{\cite{Bopp:1986} may have detected it
  earlier. One of their spectra from 1982 showed a strongly redshifted
  H$\alpha$ line in emission, very near the velocity we now expect for
  the secondary, along with absorption lines at about the right
  velocity for the other two components. However, they dismissed this
  possibility because of anomalous behaviour of the H$\alpha$ line in
  an earlier photographic spectrum, which had the lines of the two
  main components blended.\label{foot:bopp}}

\hstar\ was placed on our target list for long-baseline
interferometric observations with the Center for High Angular
Resolution Astronomy (CHARA) Array \citep{tenBrummelaar:2016}, with the goal of resolving the
inner pair to determine the component masses. As we describe here,
those observations were successful despite the small, $\sim$1~mas
semimajor axis of the orbit. The tertiary was resolved as well.
\hstar\ has also been monitored spectroscopically for more than
40 years, as part of a large survey of several hundred stars in
the Hyades region carried out at the Center for Astrophysics (CfA).
These high-resolution spectroscopic observations
have clearly revealed the lines of the faint secondary, allowing us
to measure its radial velocity with higher precision than before.
All of this material combined has enabled us to obtain highly precise
measurements of the masses for the inner pair, as well as the
brightness of each component, and the orbital parallax. We also take
the opportunity to update the astrometric orbit of the wide pair, which is
necessary in order to avoid biasing the mass determinations.
{Hierarchical triples such as this, in which the full 3D inner and
outer orbits can be determined, as we show below, are particularly useful to study
the architectures of these systems. They can provide insight into
effects such as the Lidov-Kozai cycles, which have the ability to shape their
long-term evolution \citep[see, e.g.,][]{Toonen:2016}.
}

Our interferometric and spectroscopic observations are reported in
Section~\ref{sec:observations}, along with previously obtained
measurements of the relative position of the tertiary, made mostly
with the speckle technique. Our own
measurements of the third star are given there as well. The orbital
analysis to derive the inner and outer orbits simultaneously is
described in Section~\ref{sec:analysis}. In Section~\ref{sec:fourth}
we compile evidence for a fourth star in the system, and 
Section~\ref{sec:activity} reports on the many signs of activity in \hstar.
Stellar evolution models are then used in Section~\ref{sec:discussion} for
a comparison against the observations. Our final thoughts are given in
Section~\ref{sec:conclusions}.

\section{Observations}
\label{sec:observations}

\subsection{Interferometry with CHARA}
\label{sec:chara}

The CHARA Array is located at Mount Wilson Observatory, and combines the light of six 1m telescopes with baselines ranging from 34 to 331m \citep{tenBrummelaar:2005}. 
Each telescope is equipped with an adaptive optics system that
improves the sensitivity for fainter targets such as \hstar\
\citep{Che:2013, Anagu:2020}.
We observed the object on five nights using the MIRC-X $H$-band combiner in the low resolution Prism50 mode \citep{Anugu:2020}. On the last two nights we observed simultaneously with the MYSTIC $K$-band combiner in Prism49 mode \citep{Setterholm:2023}. MIRC-X and MYSTIC combine the light from all six telescopes (S1, S2, E1, E2, W1, and W2), providing spectrally dispersed visibilities on 15 baselines and closure phases on 20 triangles. To calibrate the interferometric transfer function, we alternated between observations of unresolved calibrator stars and the science targets. The calibrators were selected using
SearchCal\footnote{\url{https://jmmc.fr/searchcal}}. The adopted uniform disk diameters for the calibrators in the $H$ and $K$ bands (UD${_H}$ and UD${_K}$) are listed in Table~\ref{tab:calibrators}.

The MIRC-X and MYSTIC data were reduced using the standard pipeline (version 1.3.5) written in python\footnote{\url{https://gitlab.chara.gsu.edu/lebouquj/mircx\_pipeline.git}}. On each night the calibrators were calibrated against each other; no evidence of binarity was found in the calibrators based on visual inspection. The calibrated OIFITS files for \hstar\, will be available in the Optical Interferometry Database\footnote{\url{http://jmmc.fr/\raisebox{0.1em}{\tiny$\sim$\,}webmaster/jmmc-html/oidb.htm}} and the CHARA Data Archive\footnote{\url{https://www.chara.gsu.edu/observers/database}}. 

The interferometric field of view is set by the coherence length of the beam combiner ($\lambda^2/\Delta\lambda$). For the low resolution ($R = 50$) spectral mode of MIRC-X, this translates to 0\farcs05 on the longest 331m baseline and 0\farcs5 on the shortest 34m baseline. Therefore, the wide tertiary companion is within the interferometric view of the shortest baselines (produces visibility modulation), but it is resolved out on the longer baselines (incoherent light produces constant scaling factor). We followed a two step approach for modelling the CHARA data. {First, we removed the data on the shorter baselines that are most impacted by the wide companion ($B < 110$m for most
  dates, and $B < 70$m for UT~2021Nov19),} and fit a simple binary model to the data. We used the IDL adaptive grid search procedure\footnote{https://www.chara.gsu.edu/analysis-software/binary-grid-search} \citep{Schaefer:2016} to solve for the separation ($\rho$), position angle east of north ($\theta$) on the International Celestial Reference System (ICRS), and the fractional fluxes of two inner components ($f_{\rm Aa}$, $f_{\rm Ab}$). We added a scaling factor ($f_{\rm B}$) to account for incoherent flux from the outer component in the field of view. The CHARA measurements for the inner pair of \hstar\, (Ab relative to Aa) are reported in Table~\ref{fig:chara}, {together with the flux contribution of the third star, expressed as the ratio $f_{\rm B}/f_{\rm Aa}$.}

We then fit the CHARA data on all baselines using the triple model described by \citet{Schaefer:2016} that includes band-width smearing. The triple model accounts for the fast visibility modulations on the shortest baselines from the wide tertiary companion. As a starting position for the triple fit, we used our measurements of the inner pair (Table~\ref{fig:chara}) and estimated the position of the outer component based on an orbit fit to the speckle observations described in Section~\ref{sec:speckle}. We then performed a grid search where we varied the position of the outer pair over a range of $\pm$20~mas in 0.5~mas steps. At each position in the grid, we optimised the separations and flux ratios of the inner and outer pairs by performing a Levenberg–Marquardt least-squares minimisation using the IDL {\sc mpfit}\footnote{http://cow.physics.wisc.edu/\raisebox{0.1em}{\tiny$\sim$\,}craigm/idl/idl.html} routine \citep{Markwardt:2009}. The CHARA measurements of 
{the position of \hstar~B relative to Aa}
are reported in Table~\ref{tab:chara_wide}.
For the two sets of data collected on UT~2020Nov12, we did not have the E2 telescope in the first set and lost delay on the E1 telescope in the second set. Therefore, we fit the triple based on the combined data for the two sets collected that night to improve the $uv$ coverage on the sky. In all cases, the triple solution produced positions for the close pair that are consistent with the simpler binary fit.

{The apparent sizes of the stars are unresolved by our observations,
even at the longest baselines. Consequently, for}
the binary and triple fits, we adopted {fixed} stellar angular diameters of 0.171, 0.125, and 0.143~mas respectively for components Aa, Ab, and B. These were estimated from
preliminary masses for the components and radii as predicted by stellar evolution
models described later. {The precise values of the diameters have a negligible effect on the results.} During the fitting process, we also divided the wavelengths in the OIFITS files by systematic correction factors of $1.0054 \pm 0.0006$ for MIRC-X and $1.0067 \pm 0.0007$ for MYSTIC (J.\ D.\ Monnier, priv.\ comm.). On UT 2022Nov15, the S1 and S2 telescopes were offline because of technical problems. The resulting $uv$ coverage provided by the E1, E2, W1, and W2 telescopes covered only a narrow swath on the sky and was insufficient for modelling the triple system.

\setlength{\tabcolsep}{12pt}
\begin{deluxetable}{lccc}
\tablewidth{0pc}
\tablecaption{Interferometric Calibrator Stars \label{tab:calibrators}}
\tablehead{
\colhead{Star} &
\colhead{${\rm UD}_H$} &
\colhead{${\rm UD}_K$} &
\colhead{$\sigma_{\rm UD}$}
\\
\colhead{} &
\colhead{(mas)} &
\colhead{(mas)} &
\colhead{(mas)}
}
\startdata
HD 17660 & 0.3053 &  0.3069 &  0.0073 \\
HD 20150 & 0.3499 &  0.3506 &  0.0127 \\
HD 23288 & 0.2277 &  0.2282 &  0.0068 \\
HD 24702 & 0.2441 &  0.2450 &  0.0057 \\
HD 27627 & 0.2727 &  0.2740 &  0.0062 \\
HD 27808 & 0.2748 &  0.2758 &  0.0066 \\
HD 28406 & 0.2766 &  0.2775 &  0.0069 \\
HD 36667 & 0.2839 &  0.2849 &  0.0069 
\enddata

\tablecomments{Uniform disk diameters in the $H$ and $K$ bands adopted from the JMMC Stellar Diameter Catalogue \citep{Bourges:2017}.}

\end{deluxetable}
\setlength{\tabcolsep}{12pt}

\setlength{\tabcolsep}{6pt}
\begin{deluxetable*}{lcccccccccl}
\tablecaption{CHARA Measurements for the Inner Binary of \hstar \label{tab:chara}}
\tablehead{
\colhead{UT Date} &
\colhead{HJD} &
\colhead{$\tau$} &
\colhead{$\rho$} &
\colhead{$\theta$} &
\colhead{$\sigma_{\rm maj}$} &
\colhead{$\sigma_{\rm min}$} &
\colhead{$\theta_{\sigma}$} &
\colhead{$f_{\rm Ab}/f_{\rm Aa}$} &
\colhead{$f_{\rm B}/f_{\rm Aa}$} &
\colhead{Instrument}
\\
\colhead{} &
\colhead{(2,400,000+)} &
\colhead{(day)} &
\colhead{(mas)} &
\colhead{(degree)} &
\colhead{(mas)} &
\colhead{(mas)} &
\colhead{(degree)} &
\colhead{} &
\colhead{} &
\colhead{}
}
\startdata
2020Oct22  &  59144.812  &  $-$0.0470  &  0.8635  &  \phn46.39  &  0.0093  &  0.0061  &  132.79     &  0.2181  &  0.7073  &  MIRC-X ($H$ band) \\
2020Oct23  &  59145.862  &  $-$0.0470  &  0.6800  &  214.46     &  0.0033  &  0.0024  &  123.84     &  0.2540  &  0.5623  &  MIRC-X ($H$ band) \\
2020Nov12  &  59165.853  &  $-$0.0470  &  0.4241  &  273.60     &  0.0085  &  0.0054  &  123.45     &  0.2817  &  0.5184  &  MIRC-X ($H$ band) \\
2020Nov12  &  59165.931  &  $-$0.0470  &  0.3957  &  308.54     &  0.0455  &  0.0306  &  129.60     &  0.1396  &  0.4697  &  MIRC-X ($H$ band) \\
2021Nov19  &  59537.847  &  $-$0.0461  &  0.8857  &  \phn65.72  &  0.0051  &  0.0047  &  164.38     &  0.2304  &  0.3601  &  MIRC-X ($H$ band) \\
2021Nov19  &  59537.847  &  $-$0.0461  &  0.8871  &  \phn65.91  &  0.0093  &  0.0073  &  \phn67.18  &  0.2577  &  0.5138  &  MYSTIC ($K$ band) 
\enddata

\tablecomments{Column $\tau$ gives the light travel time corrections
  to be added to the dates of observation, to refer them to the centre
  of mass of the triple system.
  All angles are on the ICRS (effectively J2000).
  Formal uncertainties for the flux
  ratios $f_{\rm Ab}/f_{\rm Aa}$ and $f_{\rm B}/f_{\rm Aa}$ are not
  reported, as they are typically unrealistically small. A more
  representative value for the flux uncertainties is given by the scatter of
  the measurements (see Section~\ref{sec:discussion}).}

\end{deluxetable*}
\setlength{\tabcolsep}{6pt}

\setlength{\tabcolsep}{6pt}
\begin{deluxetable*}{lccccccl}
\tablecaption{CHARA Measurements for the Tertiary of \hstar \label{tab:chara_wide}}
\tablehead{
\colhead{UT Date} &
\colhead{HJD} &
\colhead{$\rho$} &
\colhead{$\theta$} &
\colhead{$\sigma_{\rm maj}$} &
\colhead{$\sigma_{\rm min}$} &
\colhead{$\theta_{\sigma}$} &
\colhead{Instrument}
\\
\colhead{} &
\colhead{(2,400,000+)} &
\colhead{(mas)} &
\colhead{(degree)} &
\colhead{(mas)} &
\colhead{(mas)} &
\colhead{(degree)} &
\colhead{}
}
\startdata
2020Oct22 &  59144.812  &  184.888  &  220.290  &  0.112  &  0.074  &  14.121  &  MIRC-X ($H$ band) \\
2020Oct23 &  59145.862  &  186.287  &  220.147  &  0.094  &  0.071  &  31.168  &  MIRC-X ($H$ band) \\
2020Nov12 &  59165.893  &  184.288  &  220.641  &  0.071  &  0.025  &  41.062  &  MIRC-X ($H$ band) \\
2021Nov19 &  59537.847  &  175.733  &  224.485  &  0.118  &  0.028  &  47.808  &  MIRC-X ($H$ band) \\
2021Nov19 &  59537.847  &  175.675  &  224.460  &  0.108  &  0.029  &  47.516  &  MYSTIC ($K$ band) 
\enddata

\tablecomments{The CHARA measurements of the tertiary of \hstar\ (star~B) are
made relative to the primary (star~Aa).
The meaning of the columns is similar to Table~\ref{tab:chara}.}

\end{deluxetable*}
\setlength{\tabcolsep}{6pt}

\subsection{Spectroscopic Observations}
\label{sec:spectroscopy}

Our spectroscopic observations of \hstar\ at the CfA began on New
Years day, 1980. They were gathered with three different instruments on
two telescopes, as we now describe.

Observations through 2006 December were made with two nearly identical
echelle instruments on the 1.5m Wyeth reflector at the (now closed)
Oak Ridge Observatory (Massachusetts, USA), and on the 1.5m
Tillinghast reflector at the Fred L.\ Whipple Observatory (Arizona,
USA). These instruments \citep[Digital Speedometers;][]{Latham:1992}
delivered a resolving power of $R \approx 35,\!000$, and used
intensified photon-counting Reticon detectors that recorded a single
order 45\,\AA\ wide centred at a wavelength of 5187\,\AA. The main
spectral features in this region are the lines of the \ion{Mg}{1}\,b triplet. We obtained
a total of 88 spectra with these instruments. Those collected through
the end of the 2003 observing season have signal-to-noise ratios
between 10 and 22 per resolution element of 8.5~\kms. Beginning in
2004, exposures were lengthened to look for signs of the secondary,
reaching signal-to-noise ratios of 36--60. Wavelength solutions for
these observations relied on exposures of a thorium-argon lamp taken
before and after each science exposure. The zero point of the velocity
system was monitored with observations of the sky at dusk and dawn.
Small run-to-run corrections based on them were applied as described
by \cite{Latham:1992}, to place observations from both instruments on
the same native CfA system. This system is slightly offset from the
IAU system by 0.14~\kms\ \citep{Stefanik:1999}, as determined from
observations of minor planets in the solar system. We removed this
shift by adding +0.14~\kms\ to all our raw velocities from the Digital
Speedometers.

From 2011 January until 2022 December, we continued the observations
with the Tillinghast Reflector Echelle Spectrograph
\citep[TRES;][]{Furesz:2008, Szentgyorgyi:2007} on the 1.5m telescope
in Arizona. This bench-mounted, fibre-fed instrument delivers a
resolving power of $R \approx 44,\!000$, and has a CCD detector that
records 51 echelle orders between 3800\,\AA\ and 9100\,\AA. We collected 20
spectra with signal-to-noise ratios of 41--110 per resolution element
of 6.8~\kms. During each run we used observations of IAU standard stars to
monitor changes in the velocity zero-point of TRES, and observations
of asteroids to translate the raw velocities to an absolute system, as
done with the Digital Speedometers.

The weaker exposures from the Digital Speedometers, through the end of
2003, show only the lines of the primary of the inner pair (hereafter
star~Aa) and of the tertiary (star~B).  The stronger Digital
Speedometer exposures obtained later also reveal the much fainter
lines of star~{Ab}, the secondary in the inner binary. All of the TRES
observations show the lines of the three stars.

Radial velocities (RVs) for the double-lined spectra were measured using
{\tt TODCOR} \citep{Zucker:1994}, a two-dimensional cross-correlation
technique. Those in which the three stars are visible were measured
with {\tt TRICOR} \citep{Zucker:1995}, which is an extension of {\tt
  TODCOR} to three dimensions.  The templates were taken from a
pre-computed library of calculated spectra based on model atmospheres
by R.\ L.\ Kurucz, and a line list tuned to better match real stars
\citep[see][]{Nordstrom:1994, Latham:2002}. The microturbulent
velocity in these models was set to 2~\kms, and the macroturbulent velocity
to 1~\kms. For the TRES instrument,
we used only the order centred on the \ion{Mg}{1}\,b triplet, so as to
match the spectral region of the Digital Speedometers. Experience has
shown that this is also the order that provides most of the velocity
information.

The optimal template parameters were determined by running grids of
cross-correlations on the higher-quality TRES spectra, as described by
\cite{Torres:2002}, and choosing the values in our grid nearest to
those best fits.  The effective temperatures adopted for the
templates are $T_{\rm eff} = 5000$~K and 4500~K for stars~Aa and B,
respectively. Star~{Ab} is too faint for us to determine its template
parameters in the same way. The temperature in this case was taken to
be 3750~K, consistent with the properties of that star determined later.
Surface gravities of $\log g = 4.5$ were held fixed for
all three stars, along with solar composition, which is sufficiently
close to the metallicity of the Hyades for our purposes \citep[${\rm [Fe/H]} = +0.18
  \pm 0.03$;][]{Dutra-Ferreira:2016}. For the primary
template, we used a rotational broadening $V_{\rm rot}$ of 20~\kms,
which includes macroturbulence, and for stars~Ab and B we found
non-rotating templates to provide the highest cross-correlation
values {($V_{\rm rot}$ values significantly below the TRES resolution
of 6.8~\kms\ are uncertain).}
Estimates of the temperatures and $V_{\rm rot}$ for Aa and B
to a finer resolution than the sampling of our template library were
made by interpolation, resulting in values of $T_{\rm eff} = 4990 \pm
100$~K and $4510 \pm 200$~K, with rotational velocities of $V_{\rm rot} = 18.4 \pm
1.0~\kms$ and $V_{\rm rot} < 2~\kms$, respectively. An independent estimate of
the rotation of the primary by \cite{Mermilliod:2009} gave $19.3 \pm
2.0~\kms$, in good agreement with ours.  We list our velocities in
Table~\ref{tab:rvs}, along with their formal uncertainties.

\setlength{\tabcolsep}{6pt}  
\begin{deluxetable*}{lcccccccc}[!t]
\tablewidth{0pc}
\tablecaption{CfA Radial Velocity Measurements for \hstar\ \label{tab:rvs}}
\tablehead{
\colhead{HJD} &
\colhead{$\tau$} &
\colhead{Year} &
\colhead{$RV_{\rm Aa}$} &
\colhead{$RV_{\rm Ab}$} &
\colhead{$RV_{\rm B}$} &
\colhead{Inner Phase} &
\colhead{Outer Phase} &
\colhead{Instrument}
\\
\colhead{(2,400,000+)} &
\colhead{(day)} &
\colhead{} &
\colhead{(\kms)} &
\colhead{(\kms)} &
\colhead{(\kms)} &
\colhead{} &
\colhead{} &
\colhead{}
}
\startdata
 44239.6115 & $-$0.0448 & 1979.9988  & \phs$ 58.61 \pm 1.01$  &  \nodata   &    \nodata        &  0.9920 &  0.6948 & 1 \\
 44600.7619 & $-$0.0435 & 1980.9876  & $-10.32 \pm 0.92$      &  \nodata   & $41.19 \pm 1.85$  &  0.8259 &  0.7177 & 1 \\
 44603.6784 & $-$0.0435 & 1980.9956  & \phs$ 81.32 \pm 1.63$  &  \nodata   & $37.46 \pm 3.27$  &  0.0440 &  0.7179 & 1 \\
 44627.6345 & $-$0.0434 & 1981.0612  & \phs$ 83.39 \pm 1.75$  &  \nodata   & $35.22 \pm 3.51$  &  0.0492 &  0.7194 & 1 \\
 44629.6401 & $-$0.0434 & 1981.0667  & \phs$ 13.07 \pm 0.95$  &  \nodata   & $40.77 \pm 1.91$  &  0.8869 &  0.7196 & 1 
\enddata

\tablecomments{Column $\tau$ is the light travel time correction in
the outer orbit, applied in the analysis of Section~\ref{sec:analysis}
to refer the ephemerides to the barycentre of the triple system.
The formal RV uncertainties listed here are adjusted iteratively in the orbital
analysis described later.
Phases in the inner and outer orbits were computed from
the ephemerides given in Section~\ref{sec:analysis}.
The instrument code in the last column is 1 for the Digital Speedometers
and 2 for TRES.
(This table is available in its entirety in machine-readable form).}

\end{deluxetable*}
\setlength{\tabcolsep}{6pt}  

We also used our spectra to determine the relative brightness of the
components. From the Digital Speedometer observations we obtained
light ratios of $f_{\rm Ab}/f_{\rm Aa} = 0.028 \pm 0.012$ and
$f_{\rm B}/f_{\rm Aa} = 0.110 \pm 0.010$ in the
\ion{Mg}{1}\,b order. The TRES observations resulted in
similar ratios of $f_{\rm Ab}/f_{\rm Aa} = 0.028 \pm 0.004$ and
$f_{\rm B}/f_{\rm Aa} = 0.130 \pm 0.006$. We adopt weighted
averages of $f_{\rm Ab}/f_{\rm Aa} = 0.028 \pm 0.004$ and
$f_{\rm B}/f_{\rm Aa} = 0.125 \pm 0.010$. {These are all
considerably lower than the near infrared values reported above,
because the secondary and tertiary are less massive than the primary,
so their flux contribution in the optical is smaller.}

Aside from our own RVs, the only other significant set of measurements
is that of \cite{Griffin:1981}, which are the earliest for
\hstar\ (1974--1980), and can help to constrain the outer orbit. We
therefore included these 41 RVs of star~Aa in the analysis of
Section~\ref{sec:analysis}, with the weights assigned by the authors
to the individual observations.

Two other short sets of measurements are available in the literature.
\cite{Mermilliod:2009} published four RVs made with the CORAVEL
spectrometer at the Haute-Provence Observatory (France), which have
similar precision as those of \cite{Griffin:1981}, and were obtained
at similar epochs (1978--1984).  Although the publication does not
identify which component they refer to, one of the measurements, on
HJD~2,445,376.347, happens to be of star~B, and the other three are of
star~Aa. We have incorporated these observations into our analysis as
well.  Another set of five RV measurements of \hstar\ was obtained in
2004--2005 by \cite{Bender:2008}, who observed in the near infrared to
facilitate the detection of the faint star~Ab. While they succeeded
(and only reported the velocities for that star), the measurements
have relative large uncertainties and are not as useful for our
analysis.

\subsection{Speckle Observations}
\label{sec:speckle}

\begin{figure}
\epsscale{1.17}
\plotone{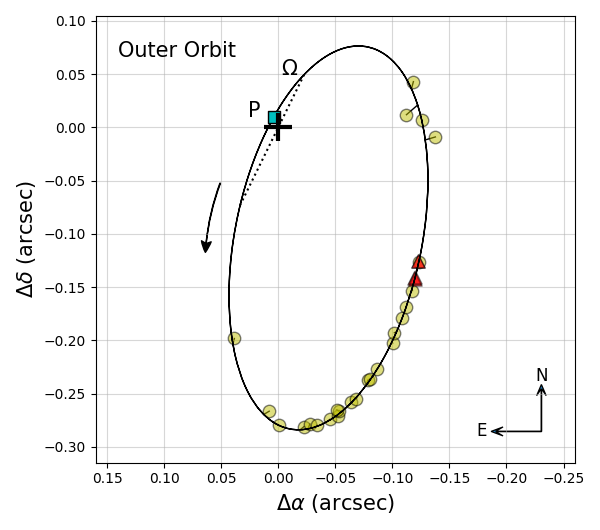}

\figcaption{Measured relative positions of the tertiary in
\hstar, together with our best fit model described in
Section~\ref{sec:analysis}. The "+" symbol represents the
centre of light of the inner binary, which is unresolved in the
speckle observations. Short line segments connect the observed
positions with the predicted location in the orbit. Circles represent
observations from the WDS, and red triangles are our own measurements
from CHARA.
The square labelled ``P'' represents
periastron, and the line of nodes is indicated with a dotted line,
where ``$\Omega$'' marks the ascending node.\label{fig:visualorbit}}

\end{figure}

\hstar\ has been followed more or less regularly by speckle observers
since the discovery of its 0\farcs14 outer companion in 1985. Some
two dozen measurements of the position of star~B relative to the
inner binary have been reported to date. A listing of these measurements
from the Washington Double Star Catalog \citep[WDS;][]{Worley:1997,
Mason:2001} was kindly provided by R.\ Matson
(USNO). The double star designation and discoverer name in the
catalogue are, respectively, WDS~J04119+2338A and CHR~14.
Two adaptive optics measurements by \cite{Morzinski:2011} and one from
\cite{Asensio-Torres:2018} were
added by us to this list, and a few minor adjustments to other WDS entries were made
after consulting the original sources. Additional measurements
from our own observations with CHARA have been given earlier in
Table~\ref{tab:chara_wide}, and are typically about an order of
magnitude more precise than the speckle measurements.
Figure~\ref{fig:visualorbit} is a graphical representation of these
observations, along with our best fit model described below.

\section{Analysis}
\label{sec:analysis}

An initial spectroscopic solution for the orbit of stars Aa and Ab
around each other
displayed an obvious pattern in the RV residuals with time, reflecting
a drift in the centre-of-mass velocity due to the motion of the inner pair
in the $\sim$40 yr outer orbit. To account for this drift properly, the
approach we have taken here is to solve for the elements of the inner and outer orbits
simultaneously, using all observations together. The radial velocities
cover an interval of 48 yr, and the speckle observations span 36 yr.
We assume here that the motions in the inner and outer orbits are
unperturbed by tidal interactions, i.e., that they are purely Keplerian.
This is justified given the large period ratio of roughly 1:6600 (see below).

The inner astrometric orbit, as mapped by the CHARA observations, is
described by the period ($P_{\rm A}$), the angular semimajor axis
($a_{\rm A}^{\prime\prime}$), the eccentricity and argument of
periastron for the secondary (via the parameters $\sqrt{e_{\rm
    A}}\cos\omega_{\rm Ab}$ and $\sqrt{e_{\rm A}}\sin\omega_{\rm
  Ab}$), the cosine of the orbital inclination angle ($\cos i_{\rm
  A}$), the position angle of the ascending node for the equinox of
2000.0 ($\Omega_{\rm A}$), and a reference time of periastron passage
($T_{\rm peri,A}$). Similar elements are used to represent the outer
orbit, indicated with the subindex ``AB''. The corresponding
argument of periastron for the tertiary in the wide orbit is $\omega_{\rm B}$. The
spectroscopic orbit of the inner binary requires two additional
parameters, $K_{\rm Aa}$ and $K_{\rm Ab}$, to represent the velocity
semiamplitudes. The radial velocity motion in the outer orbit has
similar parameters, $K_{\rm A}$ and $K_{\rm B}$. The centre-of-mass
velocity of the triple system is $\gamma$.

Not all of these elements are strictly necessary to describe the
motion of the components. This comes from the fact that the inner orbit is covered by both
spectroscopic and astrometric observations, which enables the orbital
parallax to be obtained, as follows:
\begin{equation}
\pi_{\rm orb} = \frac{2\pi}{P_{\rm A}}\frac{a_{\rm A}^{\prime\prime}
  \sin i_{\rm A}}{\sqrt{1-e_{\rm A}^2} (K_{\rm Aa}+K_{\rm Ab})}~.
\end{equation}
An analogous formula for the orbital parallax may be written using
the elements of the outer orbit. Equating those two expressions, we obtain
the following relation among the orbital elements:
\begin{equation}
K_{\rm A} + K_{\rm B} = \frac{a_{\rm AB}^{\prime\prime}}{a_{\rm
    A}^{\prime\prime}} \frac{\sin i_{\rm AB}}{\sin i_{\rm A}}
\frac{P_{\rm A}}{P_{\rm AB}} \frac{\sqrt{1-e_{\rm
      A}^2}}{\sqrt{1-e_{\rm AB}^2}} (K_{\rm Aa} + K_{\rm Ab})~.
\end{equation}
One of these elements is therefore redundant, and here we have chosen
to eliminate $K_{\rm A}$.

In Section~\ref{sec:spectroscopy} we noted the somewhat arbitrary
choice of the template for component~Ab, due to the fact that we could
not establish its parameters independently because of the faintness of
the star. This has the potential to introduce systematic errors in the
velocities of Ab, which could bias the mass determinations. To guard
against this, we introduced two additional free parameters in our
solution to represent possible offsets of the Ab velocities with
respect of those of Aa, one for the Digital Speedometer observations
($\Delta_{\rm DS}$) and another for TRES ($\Delta_{\rm TRES}$).
Similarly, we allowed for one more offset applied to the
\cite{Griffin:1981} velocities of star Aa ($\Delta_{\rm GG}$) to bring them onto
the same reference frame as the CfA observations. The CORAVEL
observations of \cite{Mermilliod:2009} are nominally on the IAU
system, as are ours, so no offset is necessary. Both \cite{Griffin:1981}
and \cite{Mermilliod:2009} reported velocities for only one component at
each epoch (usually star~Aa), ignoring the presence of the others.
While stars Ab and B are considerably fainter, in principle they can still
introduce a subtle bias in the measurements of Aa at certain phases,
which could affect its velocity semiamplitude and consequently the
masses. We would expect this effect to show up in the velocity
residuals, but as seen in a figure below, we do not detect such a
bias at a significant level compared to the uncertainties. We
therefore concluded it is safe to include these measurements of star Aa
(and B, in the case of one observation from \citealt{Mermilliod:2009}).

All in all we used 88, 14, and 87 Digital Speedometer RVs for stars
Aa, Ab, and B, respectively, 20, 18, and 20 RVs from TRES, 41 RVs of
star~Aa from \cite{Griffin:1981}, and 3 of star~Aa and one of star~B from
\cite{Mermilliod:2009}. The astrometric observations consisted of 6 pairs
of ($\theta$, $\rho$) observations
from CHARA for the inner binary, 5 pairs of CHARA measurements of the tertiary,
and 25 pairs from the WDS catalogue for the
outer orbit. The CHARA measurements for the tertiary are relative to
the primary star (Aa), and were corrected at each iteration in the
analysis to refer them to the inner binary's centre of mass. Strictly
speaking, the WDS observations of the tertiary are made relative to
the centre of light of the binary,
but the offset from the centre of mass is negligible given
the precision of those measurements.
All angles from the speckle observations have been uniformly precessed to the year 2000.0.

A joint orbital analysis of all the observations was carried out in a
Markov chain Monte Carlo (MCMC) framework, using the {\sc
  emcee}\footnote{\url{https://emcee.readthedocs.io/en/stable/index.html}}
package of \cite{Foreman-Mackey:2013}. The chains had 20,000 links
after burn-in, and we adopted uniform priors over suitable ranges for
most adjustable parameters. Convergence was checked by visual
inspection of the chains, and by requiring a Gelman-Rubin statistic of
1.05 or smaller \citep{Gelman:1992}.

Despite the best efforts by observers, it is not uncommon for
measurement errors to be either too small or too large. To ensure
proper weights for each data set, we allowed for additional free
parameters in our analysis to represent multiplicative scale factors
applied to the formal uncertainties for each kind of observation. Two
such parameters were used for the position angles and separations from
the WDS observations ($f_{\theta}$ and $f_{\rho}$), one to scale the
error ellipses of the CHARA observations of the inner binary ($f_{\rm CHARA,A}$),
and another for the tertiary measurements ($f_{\rm CHARA,AB}$).
Three additional ones were used for
the RVs of each component from the Digital Speedometers ($f_{\rm
  Aa,DS}$, $f_{\rm Ab,DS}$, $f_{\rm B,DS}$), another three for TRES
($f_{\rm Aa,TRES}$, $f_{\rm Ab,TRES}$, $f_{\rm B,TRES}$), and one more
for the \cite{Griffin:1981} velocities of star Aa ($f_{\rm Aa,GG}$).
The \cite{Mermilliod:2009} measurements were considered as part of
the TRES data set.
Log-normal priors were adopted for all of these error scaling factors. The
total number of adjustable parameters in our analysis is 32.

At each iteration we adjusted the times of observation of the RVs and
CHARA measurements to account for light travel time in the wide orbit,
following \cite{Irwin:1952, Irwin:1959}. The times of periastron
passage reported below are therefore referred to the barycentre of the
triple system.  The corrections $\tau$ range from $-0.0482$ to
$-0.0023$\,d, and are not negligible compared to the period of the
inner orbit, amounting to up to 0.02 in phase. These corrections are listed in
Table~\ref{tab:chara} and Table~\ref{tab:rvs}.

The results of our MCMC analysis are presented in
Table~\ref{tab:standard}, where for the benefit of the reader we list
the elements for the inner and outer orbits in their standard form,
rather than the form in which some of them were used in the interest
of numerical efficiency (see above).
Full results for the 32 adjustable parameters, including the adopted
priors, are given in the Appendix.
The correlations among the
elements of the inner and outer orbits from our analysis are shown
graphically also in the Appendix.

\begin{figure}
\epsscale{1.17}
\plotone{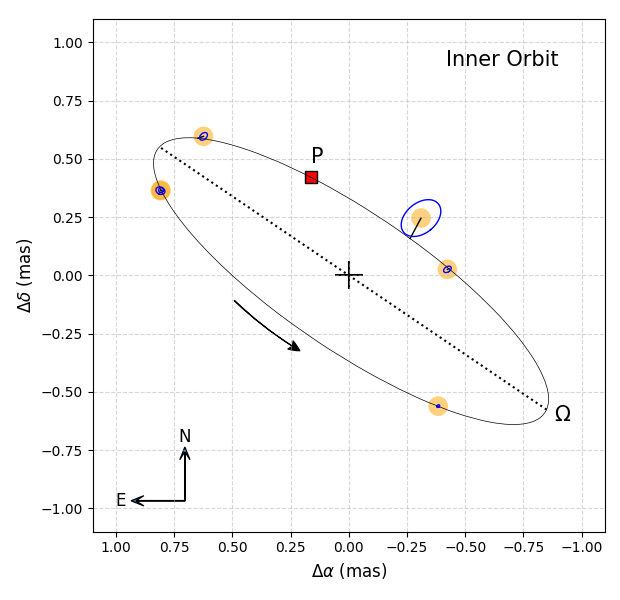}
\figcaption{CHARA observations of the inner binary in \hstar\
(highlighted with orange
circles) with their corresponding error ellipses, along with our
model for the astrometric orbit described later. The "+" sign
represents the primary (star~Aa). Short line segments connect the
observed positions with the predicted location in the orbit.
The dotted line represents the line of nodes (ascending node
labelled ``$\Omega$"), and the square labelled ``P" marks the location
of periastron.\label{fig:chara}}
\end{figure}

The inner astrometric orbit is shown in
Figure~\ref{fig:chara}, together with the CHARA observations.  The
corresponding outer orbit was shown previously in
Section~\ref{sec:speckle} (Figure~\ref{fig:visualorbit}). An earlier
astrometric solution for the outer orbit by \cite{Mason:2010}, based
on fewer measurements, has a
slightly smaller semimajor axis, and the position angle of the
ascending node in the opposite quadrant because of the
180\arcdeg\ ambiguity in some of the early speckle measurements. That
ambiguity is resolved in modern speckle observations.
A more recent orbit by \cite{Tokovinin:2023}
is much closer to ours, with all of his elements being consistent with
ours within their larger uncertainties.

\setlength{\tabcolsep}{14pt}
\begin{deluxetable}{lc}
\tablewidth{0pc}
\tablecaption{Standard Orbital Elements and Derived Properties for \hstar \label{tab:standard}}
\tablehead{
\colhead{Property} &
\colhead{Value}
}
\startdata
\multicolumn{2}{c}{Inner Orbit} \\ [0.5ex]
\hline \\ [-1.5ex]
 $P_{\rm A}$ (day)                        & $2.39436657 \pm 0.00000025$        \\ [0.5ex]
 $T_{\rm peri,A}$ (HJD)\tablenotemark{a}  & $54542.5452 \pm 0.0077$\phm{2222}  \\ [0.5ex]
 $a_{\rm A}^{\prime\prime}$ (mas)         & $1.0071 \pm 0.0044$                \\ [0.5ex]
 $e_{\rm A}$                              & $0.0557 \pm 0.0011$                \\ [0.5ex]
 $i_{\rm A}$ (deg)                        & $72.93 \pm 0.31$\phn               \\ [0.5ex]
 $\omega_{\rm Ab}$ (deg)                  & $112.9 \pm 1.2$\phn\phn            \\ [0.5ex]
 $\Omega_{\rm A}$ (deg)                   & $235.83 \pm 0.29$\phn\phn          \\ [0.5ex]
 $K_{\rm Aa}$ (\kms)                      & $66.786 \pm 0.073$\phn             \\ [0.5ex]
 $K_{\rm Ab}$ (\kms)                      & $99.88 \pm 0.55$\phn               \\ [0.5ex]
\hline \\ [-1.5ex]
\multicolumn{2}{c}{Outer Orbit} \\ [0.5ex]
\hline \\ [-1.5ex]
 $P_{\rm AB}$ (yr)                        & $43.13 \pm 0.10$\phn               \\ [0.5ex]
 $T_{\rm peri,AB}$ (yr)                   & $1993.16 \pm 0.13$\phm{222}        \\ [0.5ex]
 $a_{\rm AB}^{\prime\prime}$ (\arcsec)    & $0.3998 \pm 0.0086$                \\ [0.5ex]
 $e_{\rm AB}$                             & $0.9154 \pm 0.0039$                \\ [0.5ex]
 $i_{\rm AB}$ (deg)                       & $76.55 \pm 0.31$\phn               \\ [0.5ex]
 $\omega_{\rm B}$ (deg)                   & $78.01 \pm 0.34$\phn               \\ [0.5ex]
 $\Omega_{\rm AB}$ (deg)                  & $335.39 \pm 0.46$\phn\phn          \\ [0.5ex]
 $K_{\rm A}$ (\kms)                       & $7.8 \pm 2.1$                      \\ [0.5ex]
 $K_{\rm B}$ (\kms)                       & $17.63 \pm 0.81$\phn               \\ [0.5ex]
 $\gamma$ (\kms)                          & $+37.620 \pm 0.046$\phn\phs        \\ [0.5ex]
\hline \\ [-1.5ex]
\multicolumn{2}{c}{Derived Properties} \\ [0.5ex]
\hline \\ [-1.5ex]
 $a_{\rm A}$ ($R_{\odot}$)                 & $8.239 \pm 0.032$                 \\ [0.5ex]
 $a_{\rm AB}$ (au)                         & $15.21 \pm 0.32$\phn              \\ [0.5ex]
 $M_{\rm Aa}$ ($M_{\odot}$)                & $0.784 \pm 0.011$                 \\ [0.5ex]
 $M_{\rm Ab}$ ($M_{\odot}$)                & $0.5245 \pm 0.0047$               \\ [0.5ex]
 $q_{\rm A} \equiv M_{\rm Ab}/M_{\rm Aa}$  & $0.6686 \pm 0.0036$               \\ [0.5ex]
 $M_{\rm A}$ ($M_{\odot}$)                 & $1.309 \pm 0.015$                 \\ [0.5ex]
 $M_{\rm B}$ ($M_{\odot}$)                 & $0.59 \pm 0.12$                   \\ [0.5ex]
 $q_{\rm AB} \equiv M_{\rm B}/M_{\rm A}$   & $0.448 \pm 0.089$                 \\ [0.5ex]
 $\pi_{\rm orb}$ (mas)                     & $26.28 \pm 0.16$\phn              \\ [0.5ex]
 Distance (pc)                             & $38.05 \pm 0.24$\phn              \\ [0.5ex]
 $i_{\rm rel}$ (deg)                       & $94.94 \pm 0.53$\phn              
\enddata

\tablecomments{Values represent the mode of the corresponding
  posterior distributions, which for the derived properties and some
  of the standard elements are constructed from those of the
  quantities involved from Table~\ref{tab:results} in the Appendix.}

\end{deluxetable}
\setlength{\tabcolsep}{6pt}

Properties of the system derived from the orbital elements, such as
the masses and orbital parallax, are given at the bottom of Table~\ref{tab:standard}.
Also included is the true inclination angle between the inner and
outer orbital planes. Interestingly, the planes happen to be nearly
perpendicular to each other ($i_{\rm rel} = 94\fdg94 \pm 0\fdg53$),
formally representing retrograde motion. The eccentricity of the inner
orbit is small, but clearly not zero. 
Tidal theory predicts that the orbit of stars like these with
convective envelopes should be circularised on a timescale of
the order of 120~Myr \citep[e.g.,][]{Hilditch:2001}, which is
significantly shorter than the age of the cluster
\citep[$\sim$750~Myr;][]{Brandt:2015}. It is possible that the current
eccentricity is being maintained by the presence of the tertiary in
the system \citep[see, e.g.,][]{Mazeh:1990}.

The orbital parallax we infer
is $\pi_{\rm orb} = 26.28 \pm 0.16$~mas, compared to the value
of $27.34 \pm 0.39$~mas in the Gaia DR3 catalogue \citep[][source
 identifier 149767987810042624]{GaiaDR3:2023}. However, the presence of
the tertiary in \hstar\ appears to have degraded the quality of Gaia's
astrometric solution, as evidenced by a very large value of the
renormalised unit weight error (RUWE).\footnote{For many of the
multiple systems observed by Gaia, the processing for the DR3 catalogue
included the derivation of astrometric or spectroscopic
orbital solutions that generally improved the precision and accuracy
of the astrometric results. Unfortunately,
\hstar\ was not one of those systems with a derived orbit, likely because of the complicated
nature of the source.} Gaia sources with good-quality
solutions have RUWE values typically smaller than 1.4, with a mean of
1.0, whereas \hstar\ has 22.4.
This is not surprising, as the DR3 catalogue reports that the
stellar profiles were double-peaked almost half of the time. This implies that the
tertiary component was sometimes resolved, although the object was
still analysed as single source.
While this does not necessarily mean the Gaia parallax is incorrect,
at the very least its uncertainty will be underestimated
\citep[see][]{El-Badry:2021}.  Following these authors, and correcting
also for a zeropoint offset according to \cite{Lindegren:2021}, we
arrive at a final adjusted Gaia parallax of $27.4 \pm 1.1$~mas,
consistent with our much more precise value.

\begin{figure}
\epsscale{1.17}
\plotone{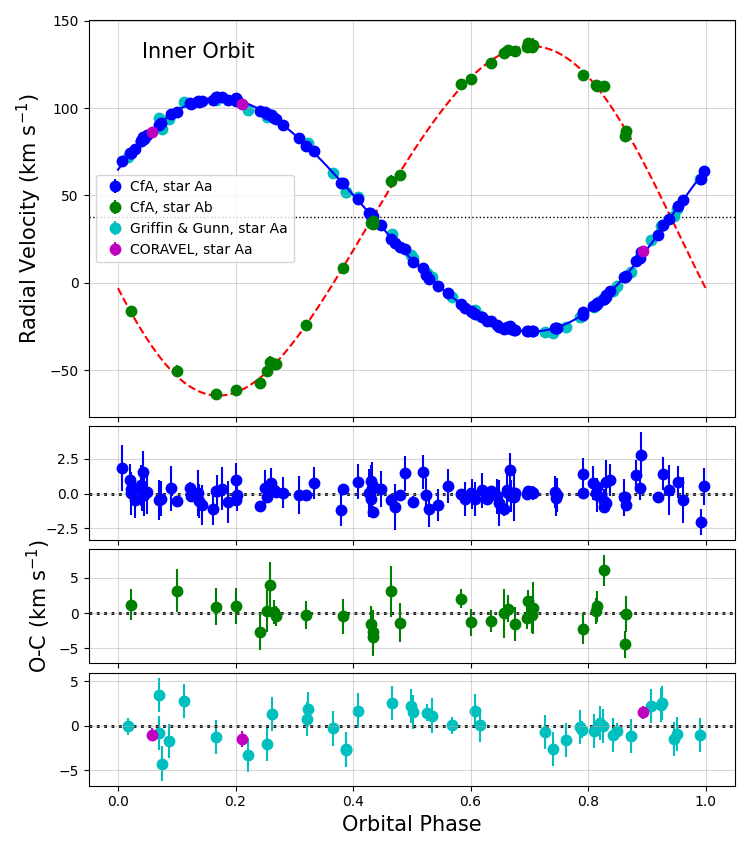}
\figcaption{Radial-velocity measurements for \hstar\ in the inner
orbit, as labelled, with motion in the outer orbit subtracted out.
The dotted line marks the centre of mass velocity of the triple system.
Residuals are shown at the bottom, separately for the CfA measurements
of stars~Aa and Ab, and other measurements of the primary.\label{fig:SBinner}}
\end{figure}

\begin{figure}
\epsscale{1.17}
\plotone{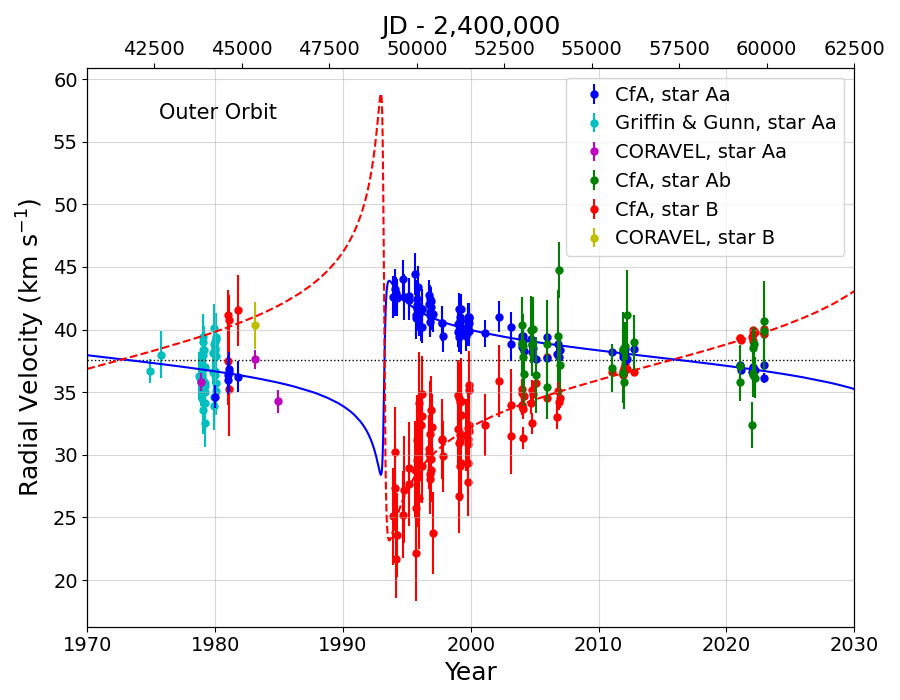}
\figcaption{Radial-velocity measurements for \hstar\ in the outer
orbit, after removing the motion in the inner orbit. The
same colour scheme is used as in Figure~\ref{fig:SBinner}.
The dotted line marks the centre-of-mass velocity of the triple system.
\label{fig:SBouter}}
\end{figure}

In Figure~\ref{fig:SBinner} we display the radial velocities for the
inner binary after subtracting the motion in the outer orbit.
Figure~\ref{fig:SBouter} shows the complementary plot of RVs in the
outer orbit, with motion in the 2.39\,d binary removed. The outer
orbit is very eccentric ($e_{\rm AB} \approx 0.92$), and the
spectroscopic observations unfortunately miss the periastron passage
of 1993, which would have more strongly constrained the velocity
amplitudes of the AB pair, strengthening the determination of the
tertiary mass. Because the tertiary's orbit is so eccentric, the
closest approach to the inner binary is only about 33 times the
size of the inner orbit.

\section{A Fourth Star in the \hstar\ System}
\label{sec:fourth}

The WDS catalogue contains several measurements of a faint star some
5\arcsec\ south of \hstar\,AB, which has a separate entry in the Gaia
DR3 catalogue (source identifier 149767987808179968).
The proper motion and parallax listed by Gaia for this object
are similar to those of the AB pair, suggesting physical
association. The WDS measurements show a slight increase {of $\sim$0\farcs7} in the
angular separation between 1997 and 2016, {and about a 2\arcdeg\ change in
the position angles over the same period,} possibly due to orbital
motion. This distant companion, which we refer to here as \hstar\,C, is
about 4.3 mag fainter than the primary in the Gaia $G$ band, but only about
1.3 mag fainter in $K_S$ \citep{Cutri:2003}, indicating it is a very red
star.\footnote{We note that the 2MASS catalogue indicates the NIR magnitudes of this companion may
be biased by contamination from a nearby star, presumably the primary. In
fact, the magnitudes of the primary star are also reported to be possibly
contaminated by a diffraction spike from a nearby star, presumably the
5\arcsec\ companion.\label{foot:2mass}}
Its radial velocity as reported by Gaia is $32.8 \pm 5.1~\kms$,
with the large uncertainty most likely due to its faintness rather
than intrinsic variability.

We have obtained two spectra of this star in 2004 using the CfA
Digital Speedometer on the 1.5m telescope in Arizona, with
signal-to-noise ratios of about 10 per resolution element.
Cross-correlations against a range of templates observed with the same
instrument indicate a best match against a spectrum of GJ~725\,B,
which is classified as M3.5 in the SIMBAD database and M4.0 by
\cite{Fouque:2018}.  Adopting a radial velocity of $1.19 \pm
0.30~\kms$ for GJ~725\,B from the latter source \citep[in agreement
  with $1.18 \pm 0.40~\kms$, from][]{Nidever:2002}, we obtain radial
velocities for \hstar\,C of $36.61 \pm 0.92~\kms$ on HJD
2,453,037.6904 and $36.39 \pm 0.80~\kms$ on HJD 2,453,276.9417. The
weighted average of these measurements, $36.48 \pm 0.60~\kms$, is only $1.1
\pm 0.6~\kms$ lower than the centre-of-mass velocity $\gamma$ of
\hstar\,AB in Table~\ref{tab:standard}, a difference consistent with being due
to orbital motion.

All the evidence therefore points toward physical association with the
brighter object, making \hstar\ a hierarchical quadruple system. At a distance to
the Earth of 38~pc from Table~\ref{tab:standard}, the projected linear
separation of \hstar\,C is 190~au, from which we infer an orbital period of roughly
1700~yr.

\section{Stellar activity}
\label{sec:activity}

\begin{figure}
\epsscale{1.17}
\plotone{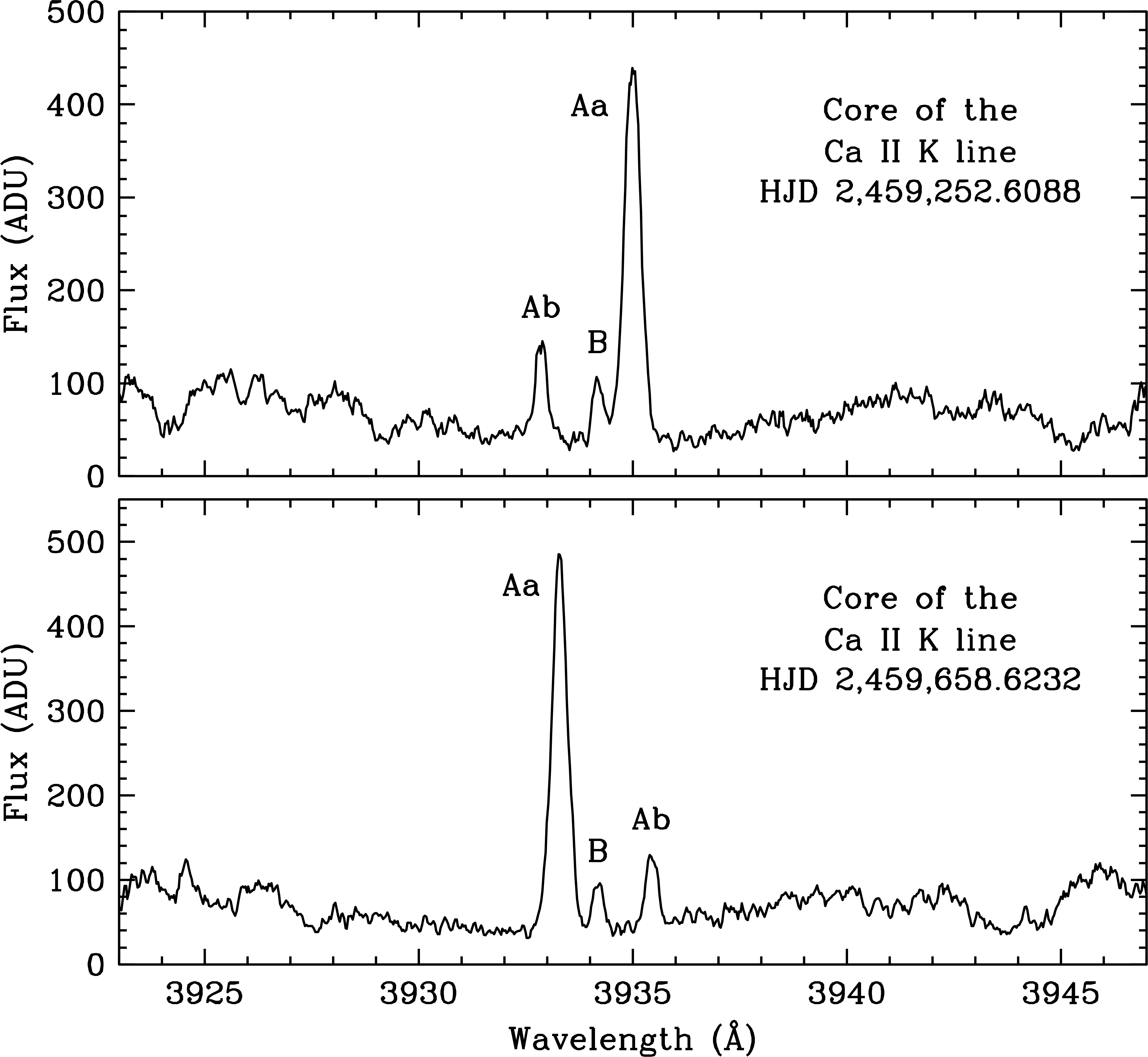}
\figcaption{Central region of the \ion{Ca}{2} K line in two of our TRES spectra,
showing that all three components in \hstar\ exhibit emission cores, indicative
of chromospheric activity. Components are labelled.
\label{fig:calcium}}
\end{figure}

\hstar\ is classified as an RS~CVn variable, and carries the designation V1136~Tau.
In addition to the brightness changes motivating its classification, which
are typically attributed to spots,
it displays other classical signatures of chromospheric activity as documented by
several authors beginning with \cite{Bopp:1986}, and more recently also by \cite{Fang:2018} and \cite{Ilin:2021}. The \ion{Ca}{2} H and K lines, H$\alpha$,
and the infrared \ion{Ca}{2} triplet are all seen in emission.
Figure~\ref{fig:calcium} shows this for the K line in our own TRES spectra, and
demonstrates that all three components of the triple system are active. In
particular, component Ab is fainter than component B, but seems to have stronger emission.
We find that H$\alpha$ is always in emission for the secondary, but not for the
tertiary (which is a slow rotator; see Section~\ref{sec:spectroscopy}), and only occasionally for the primary, though at a lower level.
This is consistent with similar findings by \cite{Bopp:1986} (see footnote~\ref{foot:bopp}). 
The system is also an extreme ultraviolet source \citep{McDonald:1994}, and an X-ray source, detected by both the ROSAT and XMM-Newton satellites.

\begin{figure*}
\epsscale{1.17}
\plotone{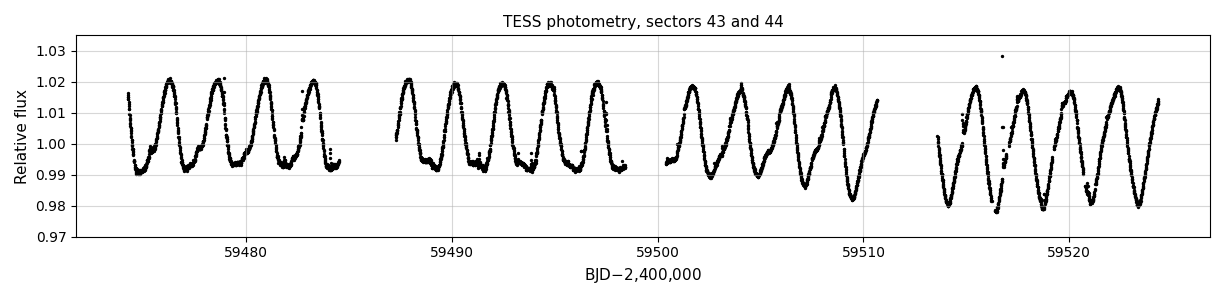}
\figcaption{Photometry of \hstar\ from the TESS mission, as recorded in the
full-frame images from sectors 43 and 44.\footnote{Observations downloaded
from the Mikulski Archive for Space Telescopes (MAST), \url{https://archive.stsci.edu}.}
Gaps in the data occur during downlinks from the satellite. 
We attribute the variability to rotational modulation, and measure a rotation period
of $P_{\rm rot} = 2.303 \pm 0.002$~d.
Numerous flares are visible on closer inspection of the original data, the largest of
which occurred on JD 2,459,516.75. It lasted for approximately 80~min, and boosted the
overall brightness of \hstar\ by 4\%. \label{fig:tess}}
\end{figure*}

Periodic brightness variations indicative of modulation by spots coming in and out of view
have been observed, but not always. \cite{Bopp:1986} searched for
variability in the early 1980s, but did not detect any significant variation.
On the other hand, the recent TESS observations show it clearly
(see Figure~\ref{fig:tess}). Rotation periods reported by various groups using different
data sets consistently give values close to, but slightly shorter than the orbital period.
\cite{Pojmanski:2002} obtained $P_{\rm rot} = 2.332$~d, with an amplitude of 0.07~mag in $V$,
based on observations from the All-Sky Automated Survey.
Photometry from the Hipparcos mission (source identifier HIP~19591) gave
$P_{\rm rot} = 2.312$~d, and an amplitude of 0.09~mag \citep{Rimoldini:2012}.
\cite{Douglas:2016} measured $P_{\rm rot} = 2.309$~d, and a total amplitude of 0.024~mag,
using photometry from the Kepler/K2 mission. From Figure~\ref{fig:tess}, we obtained
$P_{\rm rot} = 2.303 \pm 0.002$~d and a peak-to-peak amplitude of about 4\% in the TESS
band. {It is most likely that this period corresponds to the primary star,
given that the secondary is much fainter, and that the tertiary is also fainter and
much less active.}
All of these rotation period estimates are somewhat shorter than the
pseudo-synchronous period \citep{Hut:1981} {for the primary},
which is $P_{\rm pseudo} = 2.35$~d.

The photometric variability could explain some of the scatter we see in the CHARA
flux ratios reported in Table~\ref{tab:chara}. However, the differences in the table from
epoch to epoch seem larger than can be expected from changes of just 5--10\% in the
brightness of the primary, if we attribute the signal in Figure~\ref{fig:tess} to that
component alone. It seems likely, therefore, that the other components are variable as well, perhaps at a higher level than the primary {percentage-wise, even though this may not be
apparent in the light curve because of dilution from the brighter primary}.
Measurement errors may also contribute to the scatter.

\section{Discussion}
\label{sec:discussion}

Our dynamical mass determinations for the primary and secondary of \hstar\ are among the best in the Hyades (relative errors of 1.4\% and 0.9\%, respectively). The tertiary mass, on the other hand, is poorly determined mainly because of a lack of spectroscopic coverage of the outer orbit. The secondary is also the star with the lowest measured mass in the cluster to date, providing a useful extension of the empirical mass-luminosity relation toward the lower end. 

\begin{figure}
\epsscale{1.17}
\plotone{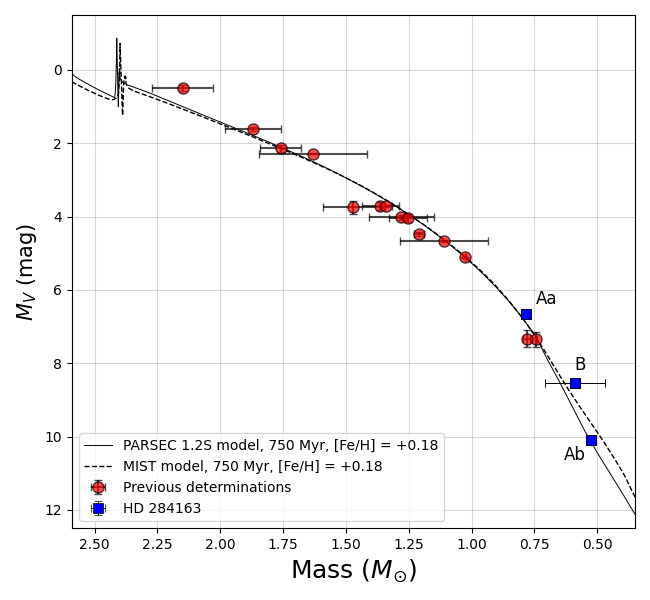}
\figcaption{Empirical mass-luminosity relation in the Hyades, for the visual band. The observations are compared against two different model isochrones, as labelled, for the age and metallicity indicated \citep[][respectively]{Brandt:2015, Dutra-Ferreira:2016}.\label{fig:mlrV}}
\end{figure}

This relation is shown in Figure~\ref{fig:mlrV} for the visual band. Masses and absolute magnitudes for the 7 binary and multiple systems in the Hyades reported prior to this work are taken from \cite{Torres:2019} and references therein, with updates for two of them \citep{Brogaard:2021, Anguita-Aguero:2022}. A complication for \hstar\ is that the overall brightness is somewhat uncertain because of the intrinsic variability. Reported $V$-band magnitudes differ by up to about 0.1~mag, ranging from $V = 9.342 \pm 0.004$ \citep{Bopp:1986} to $V = 9.445 \pm 0.028$ (from the Tycho-2 magnitudes converted to the Johnson system; \citealt{Hog:2000}). Changes in the average brightness were already noticed by \cite{Bopp:1986}, who speculated that they could be related to long-term activity cycles in \hstar, which are seen in other RS~CVn systems. Here we have chosen to adopt the brightest magnitude, on the assumption that all fainter values are impacted by spots. To deconvolve the combined light of the three stars, we used our spectroscopic flux ratios from Section~\ref{sec:spectroscopy}, transformed to the $V$ band. We did this with the aid of synthetic spectra appropriate for each star based on PHOENIX models by \cite{Husser:2013}, and interpolating the flux ratios between the \ion{Mg}{1}\,b order and the $H$ band. We obtained $f_{\rm Ab}/f_{\rm Aa}(V) = 0.043 \pm 0.004$ and $f_{\rm B}/f_{\rm Aa}(V) = 0.179 \pm 0.010$. \hstar\ was then placed on 
Figure~\ref{fig:mlrV} using our orbital parallax from Table~\ref{tab:standard}, and ignoring extinction.

Two model isochrones are shown in the figure, one from the PARSEC\,v1.2S series of \cite{Chen:2014}, and the other from the MIST series \citep{Choi:2016}. The cluster age of 750~Myr and metallicity ${\rm [Fe/H]} = +0.18$ adopted for this comparison are those proposed by \cite{Brandt:2015} and \cite{Dutra-Ferreira:2016}, respectively. The models generally fit the observations well, although formally the primary of \hstar\ is somewhat brighter than predicted. The most important difference between these series of models is in the lower main sequence, where the PARSEC\,v1.2S models have been tuned by the theorists to better match the observations of low-mass stars, by artificially altering the temperature-opacity relation below masses of about $0.75~M_{\sun}$. This change is seen to work well for the secondary of \hstar, whereas the standard MIST isochrone overpredicts its brightness.

\begin{figure}
\epsscale{1.17}
\plotone{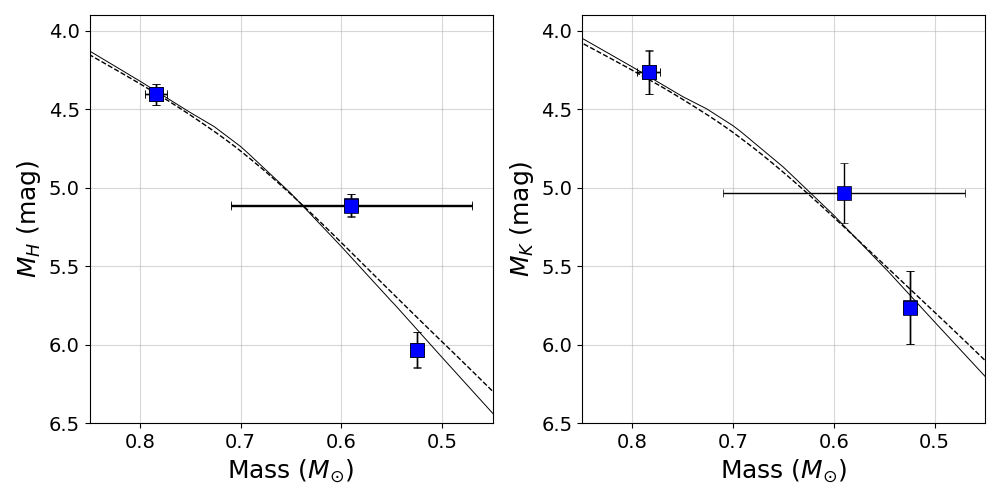}
\figcaption{$H$- and $K$-band measurements for \hstar\ shown against the same models as in Figure~\ref{fig:mlrV}.\label{fig:mlrHK}}
\end{figure}

The models fare better when compared against the absolute magnitudes in the near infrared. We computed these using the average of the $H$-band flux ratios in Table~\ref{tab:chara},
and the single values available in $K$. For the latter, we adopted uncertainties equal to
the standard deviation of the measurements in $H$. The adopted near-infrared flux ratios
are $f_{\rm Ab}/f_{\rm Aa}(H) = 0.225 \pm 0.024$, $f_{\rm B}/f_{\rm Aa}(H) = 0.524 \pm 0.057$, $f_{\rm Ab}/f_{\rm Aa}(K) = 0.258 \pm 0.053$, and $f_{\rm B}/f_{\rm Aa}(H) = 0.51 \pm 0.13$. Figure~\ref{fig:mlrHK} shows the comparison for \hstar. While both models agree in the $K$ band with the measurements for the primary and secondary, in the $H$ band only the PARSEC model fits the secondary within its error bar. None of the systems with previously published mass determinations can be shown in these diagrams, as the flux ratios and therefore the individual magnitudes are not known in the near infrared.

\begin{figure}
\epsscale{1.17}
\plotone{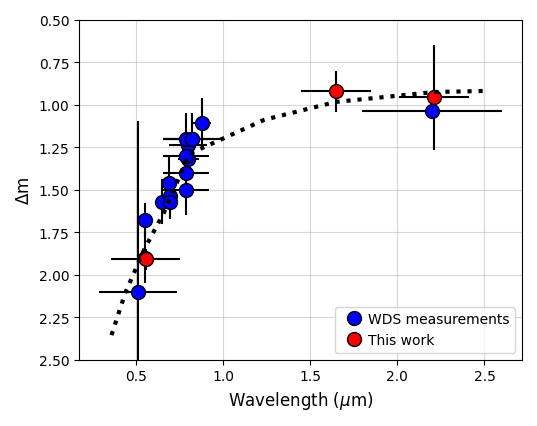}
\figcaption{Magnitude difference between the tertiary (star~B)
and the combined light of the inner binary (Aa+Ab) of \hstar, plotted as a function
of wavelength. The horizontal error bars represent the width of
each bandpass. The $\Delta m$ estimates from our
own spectroscopic and interferometric observations
agree well with the trend from other determinations in the WDS, for which we adopted
uncertainties of 10\% when not reported. The dotted
line indicates the predicted wavelength dependence of
$\Delta m$ from the PARSEC~v1.2S models
of \cite{Chen:2014}. This is based on our measured masses for the
primary and secondary, and a slightly adjusted mass for the tertiary,
which is the most uncertain of the three. We find a good match between the model and the
observations for a tertiary mass of $0.64~M_{\sun}$, which is well within the
formal uncertainty of the measured value ($M_{\rm B} = 0.59 \pm 0.12~M_{\sun}$; 
Table~\ref{tab:standard}).
\label{fig:dmag}}
\end{figure}

As a consistency check, we used our flux ratios in $V$, $H$, and $K$ to
compute the magnitude difference between the tertiary and the combined light
of the inner binary, for comparison with independent measurements at different
wavelengths as listed in the WDS. This is shown in Figure~\ref{fig:dmag}.
Our measurements agree with the run of other empirical determinations,
and follow the general trend predicted by stellar evolution models.

Finally,  it is of interest to note that our orbital parallax places \hstar\
considerably closer to us (38~pc) than the cluster centre
\citep[47.5~pc;][]{GaiaDR2:2018}. This makes a difference of about 0.5~mag in the apparent brightness. The linear distance of \hstar\ from the centre of the cluster is about 11.2~pc. This is slightly outside of the $\sim$9~pc tidal radius of the Hyades \citep[e.g.,][]{Roser:2011, Jerabkova:2021}, where stars become more strongly influenced by the Galactic potential.

\section{Conclusions}
\label{sec:conclusions}

Our CHARA observations have allowed us to spatially resolve the
inner 2.39~d binary in \hstar, a very active RS~CVn member of the Hyades
cluster. The semimajor axis we measure is only 1~mas. 
A third component in the system, previously known from
speckle observations, was also detected and
has an updated orbital period of 43.1~yr. When combined with our
RV measurements for the three stars, and others from the literature,
we have established the 3D orbits of the components, and measured
their dynamical masses, along with the orbital parallax. The inner and
outer orbital planes are nearly at right angles to each other. The masses
for the inner pair are good to better than 1.4\%, which are among
the most precise in the cluster. Our observations have also provided measures of the
relative brightness of the components, at optical as well as near-infrared
wavelengths. A fourth, fainter component about 5\arcsec\ south is also known, and
evidence is presented here that suggests it is physically associated, making
\hstar\ a quadruple system.

This is the eighth multiple system in the Hyades with mass
determinations. Comparison of the masses and absolute magnitudes with
current stellar evolution models
indicates a somewhat better fit in the near-infrared than in the optical,
and that the PARSEC~v1.2S models of \cite{Chen:2014} match the observations
of the low-mass secondary better than the standard MIST models \cite{Choi:2016}.

\section{Acknowledgements}

The spectroscopic observations of \hstar\ at the CfA were obtained
with the assistance of 
P.\ Berlind,
M.\ Calkins,
J.\ Caruso, 
G.\ Esquerdo,
E.\ Horine,
J.\ Peters, and
J.\ Zajac. 
We thank them all.
We also thank R.\ J.\ Davis and J.\ Mink for maintaining the databases of
echelle spectra, {and the anonymous referee for helpful comments.}
This work is based upon observations obtained with the Georgia State
University Center for High Angular Resolution Astronomy Array at Mount
Wilson Observatory. The CHARA array is supported by the National
Science Foundation under grant No.\ AST-1636624 and AST-2034336.
Institutional support has been provided from the GSU College of Arts
and Sciences and the GSU Office of the Vice President for Research and
Economic Development. MIRC-X received funding from the European
Research Council (ERC) under the European Union's Horizon 2020 research and
innovation program (grant No.\ 639889). 
We thank J.-B.\ Le Bouquin for contributions to the MIRC-X/MYSTIC
hardware and pipeline. J.D.M.\ acknowledges funding
for the development of MIRC-X (NASA-XRP NNX16AD43G, NSF-AST 1909165)
and MYSTIC (NSF-ATI 1506540, NSF-AST 1909165). Time at the CHARA array
was granted through the NOIRLab community access program (NOIRLab
PropID: 2020B-0010, 2021B-0008, 2022B-235883; PI: G.\ Torres). This research has
made use of the Jean-Marie Mariotti Center Aspro and SearchCal
services. S.K.\ acknowledges support by the European
Research Council (ERC Starting grant, No.\ 639889 and ERC Consolidator
grant, No.\ 101003096), and STFC Consolidated Grant (ST/V000721/1).
A.L.\ received funding from STFC studentship No.\ 630008203.

The research has made use of the SIMBAD and VizieR databases, operated
at the CDS, Strasbourg, France, of NASA's Astrophysics Data System
Abstract Service, and of the Washington Double Star Catalog maintained
at the U.S.\ Naval Observatory. We also used the WEBDA database operated
at the Department of Theoretical Physics and Astrophysics of the Masaryk
University (Czech Republic).

The work has additionally made use of data from the European Space Agency
(ESA) mission Gaia (\url{https://www.cosmos.esa.int/gaia}), processed
by the Gaia Data Processing and Analysis Consortium (DPAC,
\url{https://www.cosmos.esa.int/web/gaia/dpac/consortium}). Funding
for the DPAC has been provided by national institutions, in particular
the institutions participating in the Gaia Multilateral Agreement. The
computational resources used for this research include the Smithsonian
High Performance Cluster (SI/HPC), Smithsonian Institution
(\url{https://doi.org/10.25572/SIHPC}).


\section{Data Availability}

The data underlying this article are available in the article and in
its online supplementary material.

\appendix
\label{sec:appendix}

This appendix provides the details of our MCMC analysis.
Table~\ref{tab:results} gives the results for the 32 free parameters,
listed in the form in which they were adjusted (see the main text
in Section~\ref{sec:analysis}.

\begin{figure}
\epsscale{1.15}
\plotone{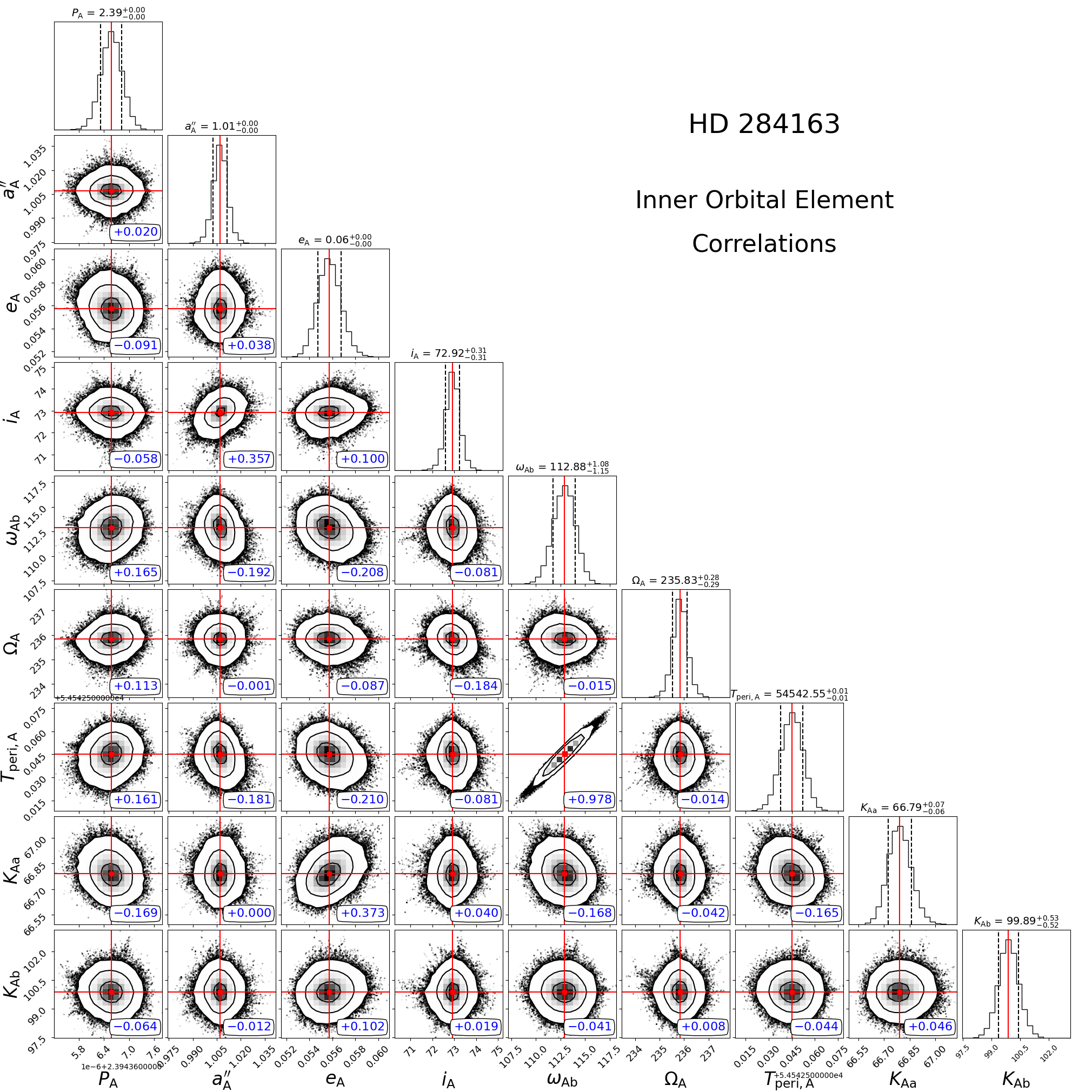}
\figcaption{Correlations among the elements of the inner orbit
of \hstar. The contours represent the 1, 2, and 3$\sigma$ confidence
levels. The correlation coefficients are indicated
in each panel.\label{fig:inner_correlations}}
\end{figure}

\begin{figure}
\epsscale{1.15}
\plotone{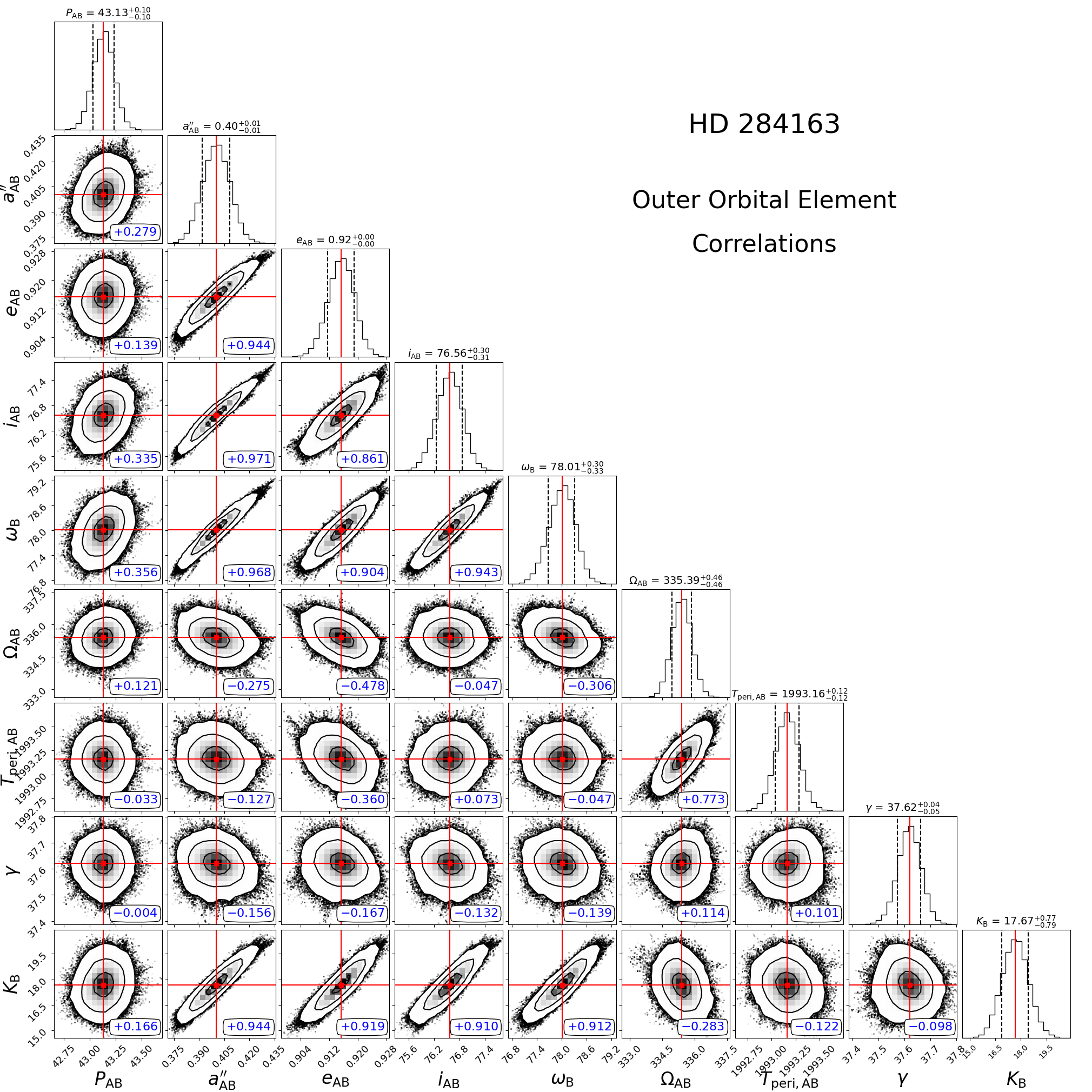}
\figcaption{Correlations among the elements of the outer orbit
of \hstar. The contours represent the 1, 2, and 3$\sigma$ confidence
levels. The correlation coefficients are indicated
in each panel.\label{fig:outer_correlations}}
\end{figure}

Figures~\ref{fig:inner_correlations} and \ref{fig:outer_correlations}
show the correlations among the
elements of the inner and outer orbits of \hstar, respectively,
based on the posterior distributions from our MCMC analysis.
The elements of the outer orbit display higher correlations,
as expected from the more incomplete observational coverage.

\setlength{\tabcolsep}{6pt}
\begin{deluxetable}{lcc}
\tablewidth{0pc}
\tablecaption{Full Set of Adjusted Parameters for \hstar \label{tab:results}}
\tablehead{
\colhead{Parameter} &
\colhead{Value} &
\colhead{Prior}
}
\startdata
\multicolumn{3}{c}{Inner Orbit} \\ [0.5ex]
\hline \\ [-1.5ex]
 $P_{\rm A}$ (day)                        & $2.39436657 \pm 0.00000025$        & [2, 3]           \\ [0.5ex]
 $T_{\rm peri,A}$ (HJD)\tablenotemark{a}  & $54542.5452 \pm 0.0077$\phm{2222}  & [54541, 54543]   \\ [0.5ex]
 $a_{\rm A}^{\prime\prime}$ (mas)         & $1.0071 \pm 0.0044$                & [0.5, 2.0]       \\ [0.5ex]
 $\sqrt{e_{\rm A}}\cos\omega_{\rm Ab}$    & $-0.0918 \pm 0.0043$\phs           & [$-1$, 1]        \\ [0.5ex]
 $\sqrt{e_{\rm A}}\sin\omega_{\rm Ab}$    & $+0.2174 \pm 0.0031$\phs           & [$-1$, 1]        \\ [0.5ex]
 $\cos i_{\rm A}$                         & $0.2936 \pm 0.0052$                & [$-1$, 1]        \\ [0.5ex]
 $\Omega_{\rm A}$ (deg)                   & $235.83 \pm 0.29$\phn\phn          & [0, 360]         \\ [0.5ex]
 $K_{\rm Aa}$ (\kms)                      & $66.786 \pm 0.073$\phn             & [50, 150]        \\ [0.5ex]
 $K_{\rm Ab}$ (\kms)                      & $99.88 \pm 0.55$\phn               & [50, 150]        \\ [0.5ex]
\hline \\ [-1.5ex]
\multicolumn{3}{c}{Outer Orbit} \\ [0.5ex]
\hline \\ [-1.5ex]
 $P_{\rm AB}$ (yr)                        & $43.13 \pm 0.10$\phn               & [30, 50]         \\ [0.5ex]
 $T_{\rm peri,AB}$ (yr)                   & $1993.16 \pm 0.13$\phm{222}        & [1980, 2010]     \\ [0.5ex]
 $a_{\rm AB}^{\prime\prime}$ (\arcsec)    & $0.3998 \pm 0.0086$                & [0.1, 0.6]       \\ [0.5ex]
 $\sqrt{e_{\rm AB}}\cos\omega_{\rm B}$    & $+0.1987 \pm 0.0051$\phs           & [$-1$, 1]        \\ [0.5ex]
 $\sqrt{e_{\rm AB}}\sin\omega_{\rm B}$    & $+0.9359 \pm 0.0031$\phs           & [$-1$, 1]        \\ [0.5ex]
 $\cos i_{\rm AB}$                        & $0.2325 \pm 0.0053$                & [$-1$, 1]        \\ [0.5ex]
 $\Omega_{\rm AB}$ (deg)                  & $335.39 \pm 0.46$\phn\phn          & [0, 360]         \\ [0.5ex]
 $K_{\rm B}$ (\kms)                       & $17.63 \pm 0.81$\phn               & [0, 50]          \\ [0.5ex]
\hline \\ [-1.5ex]
\multicolumn{3}{c}{Other Spectroscopic Parameters} \\ [0.5ex]
\hline \\ [-1.5ex]
 $\gamma$ (\kms)                          & $+37.620 \pm 0.046$\phn\phs        & [20, 50]         \\ [0.5ex]
 $\Delta_{\rm DS}$ (\kms)                 & $-0.32 \pm 0.73$\phs               & [$-10$, 10]      \\ [0.5ex]
 $\Delta_{\rm TRES}$ (\kms)               & $+1.35 \pm 0.52$\phs               & [$-10$, 10]      \\ [0.5ex]
 $\Delta_{\rm GG}$ (\kms)                 & $+0.10 \pm 0.27$\phs               & [$-10$, 10]      \\ [0.5ex]
\hline \\ [-1.5ex]
\multicolumn{3}{c}{Astrometric Error Inflation Factors} \\ [0.5ex]
\hline \\ [-1.5ex]
 $f_{\theta}$                             & $2.24 \pm 0.41$                    & [$-5$, 5]        \\ [0.5ex]
 $f_{\rho}$                               & $2.69 \pm 0.53$                    & [$-5$, 5]        \\ [0.5ex]
 $f_{\rm CHARA,A}$                        & $1.44 \pm 0.50$                    & [$-5$, 5]        \\ [0.5ex]
 $f_{\rm CHARA,AB}$                       & $5.8 \pm 2.2$                      & [$-5$, 5]        \\ [0.5ex]
\hline \\ [-1.5ex]
\multicolumn{3}{c}{Spectroscopic Error Inflation Factors} \\ [0.5ex]
\hline \\ [-1.5ex]
 $f_{\rm Aa,DS}$                          & $0.914 \pm 0.080$                  & [$-5$, 5]        \\ [0.5ex]
 $f_{\rm Ab,DS}$                          & $1.32 \pm 0.34$                    & [$-5$, 5]        \\ [0.5ex]
 $f_{\rm B,DS}$                           & $1.062 \pm 0.093$                  & [$-5$, 5]        \\ [0.5ex]
 $f_{\rm Aa,TRES}$                        & $1.32 \pm 0.30$                    & [$-5$, 5]        \\ [0.5ex]
 $f_{\rm Ab,TRES}$                        & $0.99 \pm 0.23$                    & [$-5$, 5]        \\ [0.5ex]
 $f_{\rm B,TRES}$                         & $1.05 \pm 0.25$                    & [$-5$, 5]        \\ [0.5ex]
 $f_{\rm Aa,GG}$                          & $1.06 \pm 0.14$                    & [$-5$, 5]  
\enddata

\tablenotetext{a}{The time of periastron passage is referenced to HJD~2,400,000.}
\tablecomments{Values listed correspond to the mode of the respective
  posterior distributions, with uncertainties representing the 68.3\%
  credible intervals. Priors in square brackets are uniform over the
  ranges specified, except those for the error inflation factors $f$,
  which are log-uniform.}

\end{deluxetable}
\setlength{\tabcolsep}{6pt}

\end{document}